\documentclass[]{article}

\usepackage[affil-it]{authblk}
\usepackage[utf8]{inputenc}
\usepackage{amsmath}
\usepackage{amssymb}
\usepackage{amsthm}
\usepackage{graphicx}
\usepackage{booktabs}
\usepackage{hyperref}
\usepackage[capitalize]{cleveref}
\usepackage[round]{natbib}
\usepackage{geometry}
\usepackage{url}
\usepackage{lineno}

\newtheorem{thm}{Theorem} 
\newtheorem{lemma}[thm]{Lemma} 
\newtheorem{corollary}[thm]{Corollary}
\newtheorem{proposition}[thm]{Proposition}

\title{The inviscid Euler limit as a critical boundary for moment-based aerodynamic system identification}
\author{Sarasija Sudharsan}
\affil{\small Department of Aerospace Engineering, Embry-Riddle Aeronautical University, Prescott, AZ, 86301.}
\date{\today}

\begin{document}
\maketitle

\begin{abstract}
Finite-dimensional state-space representations of unsteady aerodynamics implicitly assume a system with fading memory. 
However, the impulse response of the two-dimensional inviscid (Euler) equations is characterized by an asymptotic $t^{-3/2}$ power-law decay due to the persistence of shed vorticity.
The present work demonstrates that this decay rate constitutes a critical boundary for moment convergence: the second temporal moment diverges logarithmically, causing the characteristic memory time to grow as $\sqrt{\ln T}$ with the observation window $T$.
As a result, no window-independent characteristic time scale exists, and finite-dimensional models fitted to inviscid data effectively parameterize the observation horizon rather than intrinsic flow physics.
To quantify this behavior, a temporal-moment diagnostic, $\nu_t(T)$, is introduced based on the ratio of the second and zeroth windowed moments of the impulse response kernel.
Exponential models exhibit stable memory time plateaus, as their sufficiently fast decay ensures convergence of the moment diagnostic.
Compressible Euler simulation results confirm the predicted $\sqrt{\ln T}$ scaling at intermediate times, while numerical dissipation inherent to the discretization acts as an artificial regularizer that enforces convergence at late times.
These results establish the two-dimensional inviscid limit as a critical boundary for moment-based system identification, where the absence of a dissipative mechanism prevents the definition of a window-independent characteristic memory time.
\end{abstract}

\section{Introduction}
Reduced-order models for unsteady aerodynamics approximate the history-dependent flow response using finite-dimensional state-space representations.
This approach implicitly assumes that the system's memory can be characterized by a finite set of intrinsic time scales.
However, the impulse response of the inviscid (Euler) equations decays as a power law scaling, namely, $t^{-3/2}$, due to the persistence of shed vorticity, rather than exponentially, as imposed by finite-dimensional models.
The present work demonstrates that this power-law decay rate sits exactly at a critical boundary for convergence. 
The second temporal moment of the impulse response kernel diverges logarithmically, causing the characteristic memory time to grow as $\sqrt{\ln T}$ with the observation window $T$.
Therefore, a window-independent finite-dimensional representation cannot be rigorously defined for this case, establishing that finite-dimensional models fitted to inviscid data effectively parameterize the observation horizon rather than intrinsic flow physics.

Early analytical developments in unsteady aerodynamics relied on potential flow theory to calculate the transient forces on a thin airfoil undergoing arbitrary motion.
Theodorsen~\citep{theodorsen1935} derived the exact frequency-domain lift response for an airfoil undergoing harmonic oscillations, while Wagner~\citep{wagner1925} formulated the time-domain indicial response (the aerodynamic lift generated for a step change in angle of attack).
These foundational theories established that the transient lift does not decay exponentially toward its steady state.
Instead, in a purely inviscid, two-dimensional flow, the absence of viscous dissipation ensures that the vorticity shed from the trailing edge into the wake persists indefinitely.
As this starting vortex convects downstream, its induced velocity acting on the airfoil decays algebraically, specifically following a $t^{-3/2}$ power law decay~\citep{sears1940, lomax1952}.
Despite this well-established power-law, modern control methods approximate aerodynamic responses using exponential functions.
Classical approximations, such as the R.T. Jones representation of the Wagner function~\citep{jones1940} (given by~\eqref{eq:jones}), enforce exponential decay and thereby embed the system within a finite-dimensional Markovian framework.
\begin{equation}
\phi(t) = 1 - c_1 e^{-\lambda_1 t} - c_2 e^{-\lambda_2 t}
\label{eq:jones}
\end{equation}
where $c_1, c_2$ are weighting coefficients and $\lambda_1, \lambda_2$ represent the empirical non-circulatory and circulatory decay rates, respectively. This structure guarantees the existence of a finite characteristic time scale, which is a prerequisite for optimal control strategies such as the Linear Quadratic Regulator (LQR)~\citep{anderson1990} and Model Predictive Control (MPC)~\citep{camacho2013}.
While computationally expedient, enforcing exponential decay fundamentally alters the asymptotic physics of the theoretical Euler limit.

System Identification (SysID) techniques, such as the Eigensystem Realization Algorithm (ERA)~\citep{juang1985} and the Observer Kalman Filter Identification (OKID)~\citep{juang1993}, fit a finite-dimensional state-space model to a given set of data.
These algorithms identify a characteristic scale, namely, a dominant frequency or decay rate, that describes the system.
Sparse identification (SINDy) has been used to capture the long-time asymptotic behavior of the Wagner function~\citep{dawson2021}, yet remains constrained to finite-dimensional dynamical representations.
While the $t^{-3/2}$ asymptotic nature of Wagner's analytical function is well-documented as a source of long-term relative error in finite-order models~\citep{dawson2021}, its implications for the window-independence of temporal metrics have been largely unexplored.

The present work reframes the $t^{-3/2}$ asymptotic scaling as not merely a source of quantitative error, but a critical boundary that precludes a window-independent system representation.
By formulating the aerodynamic memory in terms of diverging temporal moments, it is shown that moment-based characterization of inviscid flow does not converge to a window-independent result, establishing why empirical data-driven models of these systems inherently require dissipative regularization.
This specific power-law decay creates a non-vanishing bias in moment-based estimates, revealing that the characteristic time scale of an inviscid system is an observation-dependent quantity rather than a physical constant.

The mechanism is as follows: while the total energy of the kernel ($M_0$) remains finite, its temporal spread ($M_2$) diverges, precluding a window-independent representation.
For short observation windows $T$, a stable finite-dimensional model can be obtained.
As $T$ increases, progressively lower-frequency states are required to capture the $t^{-3/2}$ tail, and in the limit $T \to \infty$, no finite-dimensional model can faithfully represent the kernel.
The moment-based diagnostic $\nu_t$ developed herein quantifies this non-convergence and establishes why classical state-space models cannot yield a window-independent representation of inviscid dynamics. 
This result also poses a critical question for control-oriented modeling: if the inviscid system lacks a finite memory time scale, in what sense do finite-dimensional models derived from Euler-based data reflect intrinsic fluid dynamics versus the effects of implicit regularization, and how does this affect their generalizability?

The remainder of this paper is organized as follows.
The temporal moment metrics are first introduced, followed by the mathematical framework proving the logarithmic divergence at the critical exponent.
The physical implications for 2D and 3D identifiability are then discussed, and computational results using an inviscid Euler simulation are compared with theoretical results.

\section{Nomenclature and Metric Definitions}
Consider a linear aerodynamic system characterized by its impulse response kernel $G(t)$, relating the input motion to the resulting lift response through convolution:
\begin{equation}
    y(t) = \int_0^{\infty} G(\tau) u(t-\tau) \, d\tau
\end{equation}

where $y(t)$ is the system output and $u(t)$ is the system input.
Here, $G(t)$ represents the system’s response to an impulse input. The convergence properties of its temporal moments are examined in the following sections.

To quantify the existence of a characteristic temporal scale, windowed temporal moments are defined over a finite observation horizon $T$. The $k^{\text{th}}$ windowed temporal moment, $M_k(T)$ is given by~\eqref{eq:defmoments}.

\begin{equation}
    M_k(T) = \int_0^T t^k G(t)^2 \, dt
    \label{eq:defmoments}
\end{equation}

Two metrics are considered that provide complementary characterizations of the kernel, corresponding to the temporal spread (analogous to variance in probability theory) and the spectral Rice frequency, which are commonly used in signal processing to characterize the temporal and spectral properties of signals, respectively.

\begin{itemize}

    \item \textbf{Temporal Spread Metric ($\nu_t$):} Calculated using the second and zeroth temporal moments of the kernel. This represents the effective temporal spread of the system memory.
\begin{equation}
    \nu_t(T) = \sqrt{\frac{\int^T_0 t^2 G(t)^2 \rm dt}{\int^T_0 G(t)^2 \rm dt}} = \sqrt{\frac{M_2(T)}{M_0(T)}} 
\label{eq:defnut}
\end{equation}

where the definitions of $M_2(T)$ and  $M_0(T)$ from \eqref{eq:defmoments} are used. Although denoted $\nu_t(T)$, this quantity has dimensions of time and is, therefore, interpreted as a temporal spread scale, representing the characteristic memory time of the impulse response.

\item \textbf{Spectral Rice Frequency ($\nu_s$):} Defined analogously to the classical Rice frequency~\citep{rice1944,rice1945} as the ratio of spectral moments as given by~\eqref{eq:defricefreq}. This is equal to the ratio of the squared derivative of the kernel to the kernel energy and represents the average frequency of the oscillations in the signal.

\begin{equation}
            \nu_s(T) = \frac{1}{2\pi} \sqrt{\frac{\int^{T}_0 \omega^2 S_T(\omega) \rm d \omega}{\int^{T}_0 S_T(\omega) \rm d \omega}} =  \frac{1}{2\pi} \sqrt{\frac{\int^{T}_0 \dot{G}(t)^2 \rm dt}{\int^{T}_0 G(t)^2 \rm dt}}
        \label{eq:defricefreq}
\end{equation}
where $\omega$ is the angular frequency and $S_T(\omega)$ is the windowed power spectral density of the kernel.

\end{itemize}

\noindent The metric $\nu_t(T)$ from \eqref{eq:defnut} is interpreted as follows:
\begin{itemize}
    \item $M_0(T)$ represents the total kernel energy contained within the observation window $T$,
    \item $M_2(T)$ measures the temporal spread or distribution of that energy,
    \item $\nu_t(T)$ serves as a measure of the effective or characteristic memory time scale.
\end{itemize}

\noindent Convergence of $\nu_t(T$) in the limit $T \to \infty$ is interpreted as evidence of an intrinsic characteristic temporal scale.
Divergence of $\nu_t(T)$ indicates a scale-free memory.

\noindent In this study, $\nu_t$ is prioritized as the primary diagnostic metric, as it directly captures the divergence of temporal moments associated with power-law decay.

\section{Mathematical Framework}
The existence of a window-independent system representation is determined by the asymptotic decay of the impulse response kernel $G(t)$.
To analyze the convergence of the windowed kernel function moments, the truncation error $\Delta(T)$, applicable when the infinite-horizon limit exists, is defined as
\begin{equation}
    \Delta(T) = \nu_t(\infty) - \nu_t(T)
    \label{eq:defdelta}
\end{equation}
where the infinite-horizon limits depend on the convergence of the tail contributions, $\delta M_k(T) = \int_T^\infty t^k G^2(t) \, dt$. Different classes of kernel function tails are considered, and their convergence properties are analyzed by evaluating the truncation error.

\subsection{Case I: The Exponential Kernel}
Consider a kernel with exponential decay $G(t) = C e^{-at}$. Substituting this into the moment definition:
\begin{equation}
    M_0(\infty) = C^2 \int_0^\infty e^{-2at} \, dt = \frac{C^2}{2a}
\end{equation}
\begin{equation}
    M_2(\infty) = C^2 \int_0^\infty t^2 e^{-2at} \, dt = \frac{C^2}{4a^3}
\end{equation}
The resulting temporal spread scale $\nu_t = \frac{1}{2a}$ is strictly constant and finite for all $a > 0$, and converges to this value as $T \longrightarrow \infty$.
This represents a system with short memory, where the response is localized in time.
Since the tail error $\delta M_k(T)$ decays exponentially ($e^{-2aT}$), the truncation error $\Delta(T)$ vanishes rapidly. 
This represents the standard state-space assumption of finite memory.

\subsection{Case II: The Power-Law Kernel}
For systems characterized by long-range temporal correlations, the kernel follows a power-law decay $G(t) \sim C t^{-\alpha}$.

\subsubsection{Admissibility and Critical Threshold}
\label{sec:admissibility}
To evaluate the existence of a window-independent system representation, the convergence criteria for the temporal moments of the kernel must first be established.

\begin{lemma}[Admissibility of the $k$-th Moment]
    For an impulse response kernel $G(t)$ exhibiting asymptotic power-law decay $G(t) \sim t^{-\alpha}$ as $t \to \infty$, the $k$-th temporal moment $M_k(\infty) = \int_0^\infty t^k G^2(t) \, dt$ is admissible (finite) if and only if the decay exponent satisfies:
    \begin{equation}
        \alpha > \frac{k+1}{2}
        \end{equation}
    \end{lemma}
\begin{proof}

The integrand of $M_k(\infty)$ behaves asymptotically as $t^k (t^{-\alpha})^2 = t^{k-2\alpha}$.
The integral $\int_{t_0}^\infty t^p dt$ converges if and only if $p < -1$.
Substituting $p = k-2\alpha$ yields $k-2\alpha < -1$, which simplifies to the stated condition $\alpha > (k+1)/2$.
\end{proof}

\noindent Setting $k=2$ for the second temporal moment yields the following necessity condition:
\begin{corollary}[Existence of $\nu_t(\infty)$]
The temporal spread metric $\nu_t(\infty)$ is finite if and only if $\alpha > \frac{3}{2}$.
\end{corollary}

This result identifies a critical threshold at $\alpha = \frac{3}{2}$ separating two fundamentally different regimes.
For $\alpha > \frac{3}{2}$, the system admits a finite temporal scale and a window-independent characterization. 
For $\alpha \le \frac{3}{2}$, the second moment diverges, and the system becomes scale-free with no well-defined temporal spread scale.

\subsubsection{Convergent Power-law Regimes}
\label{sec:biaslaw}
For systems in the convergent regime ($\alpha > \frac{3}{2}$), the truncation error $\Delta(T) = \nu_t(\infty) - \nu_t(T)$ is defined to quantify the truncation bias introduced by finite observation.

\begin{proposition}[Truncation Error Scaling]
\label{prop:scaling_law}
In the convergent power-law regime, the truncation error follows:
\begin{equation}
    \Delta(T) \sim \mathcal{O}(T^{3-2\alpha})
\end{equation}
\end{proposition}

\begin{proof}
Let $M_k(\infty) = M_k(T) + \delta M_k(T)$, where $\delta M_k(T) = \int_T^\infty t^k G^2(t) dt$ represents the tail contribution. 
For $T \gg 1$:
\begin{equation}
    \delta M_k(T) \approx C^2 \int_T^\infty t^{k-2\alpha} dt = \frac{C^2}{2\alpha - (k+1)} T^{k+1-2\alpha}
\end{equation}
The difference of the squares is evaluated to avoid nonlinearities:
\begin{align}
    \nu_t^2(\infty) - \nu_t^2(T) &= \frac{M_2(\infty)}{M_0(\infty)} - \frac{M_2(\infty) - \delta M_2}{M_0(\infty) - \delta M_0} \\
    &= \frac{M_0(\infty)\delta M_2 - M_2(\infty)\delta M_0}{M_0(T)M_0(\infty)}
\end{align}

Since $\delta M_2 \sim T^{3-2\alpha}$ and $\delta M_0 \sim T^{1-2\alpha}$, the term $\delta M_2$ decays more slowly and dominates the asymptotic behavior of the numerator as $T \to \infty$. The denominator can be rewritten as $M_0(T)M_0(\infty) = (M_0(\infty) - \delta M_0) M_0(\infty)$. Neglecting higher-order terms and the vanishing $\delta M_0$ term in the denominator:
\begin{equation}
    \nu_t^2(\infty) - \nu_t^2(T) \approx \frac{\delta M_2(T)}{M_0(\infty)} \propto T^{3-2\alpha}
\end{equation}
Applying the linearization $\Delta(T) \approx \frac{\nu_t^2(\infty) - \nu_t^2(T)}{2\nu_t(\infty)}$, the scaling law $\Delta(T) \sim \mathcal{O}(T^{3-2\alpha})$ is recovered.
\end{proof}
\noindent Since $\alpha > \frac{3}{2}$, the exponent $3 - 2 \alpha < 0$, ensuring that $\Delta(T) \longrightarrow 0$ as $T \longrightarrow \infty$.

\subsection{Critical Case: Inviscid Euler Limit}
\label{sec:euler}
The theoretical inviscid (Euler) limit is characterized by a $t^{-3/2}$ power-law decay in its impulse response. 
As demonstrated by the proof in Section~\ref{sec:admissibility}, this specific decay rate sits exactly at the critical threshold $\alpha = \frac{3}{2}$ for convergence of the temporal spread scale.
The inviscid asymptotic impulse response is well-established to satisfy:
\begin{equation}
    G(t) = C t^{-3/2} + o(t^{-3/2}).
\end{equation}
While this signal possesses finite energy, its decay rate is insufficient to yield a finite second temporal moment.
In this limit, the error $\Delta(T)$ given by Eq.~\eqref{eq:defdelta} is not defined, as $\nu_t(\infty)$ diverges.

\begin{thm}[Logarithmic Divergence for Euler Decay]
If $G(t) \sim C t^{-3/2}$ as $t \to \infty$, the temporal moments and the resulting temporal spread scale exhibit the following asymptotic behavior:
\begin{enumerate}
    \item The zeroth moment $M_0(T)$ converges to a finite constant.
    \item The second moment $M_2(T)$ exhibits logarithmic divergence, $M_2(T) \sim C^2 \ln T$.
    \item The temporal spread scale $\nu_t(T)$ grows as $\sqrt{\ln T}$.
\end{enumerate}
\end{thm}

\begin{proof}
Using the asymptotic form $G^2(t) \sim C^2 t^{-3}$,
\begin{equation}
    M_0(T) = \int_{1}^{T} C^2 t^{-3} \, dt = \frac{C^2}{2}(1 - T^{-2}) \xrightarrow{T \to \infty} \frac{C^2}{2},
\end{equation}
Note that since the asymptotic form $G(t) \sim C t^{-3/2}$ is valid only for large $t$, the integral is restricted to $t \ge 1$; the contribution from $t \in [0,1]$ is finite and does not affect the convergence properties.
The result above confirms that the total energy remains bounded.
However, the second moment weights the tail by $t^2$, yielding a logarithmically divergent integral.
\begin{equation}
    M_2(T) = \int_{1}^{T} t^2 (C^2 t^{-3}) \, dt = C^2 \int_{1}^{T} t^{-1} \, dt = C^2 \ln T.
\end{equation}
Substituting these results into the temporal spread scale $\nu_t(T) = \sqrt{M_2(T)/M_0(T)}$ yields:
\begin{equation}
    \nu_t(T) \sim \sqrt{\frac{C^2 \ln T}{M_0(\infty)}} \sim \sqrt{\ln T}.
\end{equation}
\end{proof}
\noindent This logarithmic divergence represents the marginal case identified in the admissibility analysis, where the second temporal moment fails to converge but does so at the slowest possible rate.

Physically, this $t^{-3/2}$ power-law decay originates from the farfield induction of the starting vortex as it convects downstream~\citep{sears1940}.
In the 2D inviscid limit, the absence of vorticity diffusion preserves the coherent structure of the wake, leading to a persistent memory effect where the second temporal moment and thus the characteristic memory time fails to converge to a window-independent value.

\subsection{Comparative Scaling of the Spectral Rice Frequency}
While $\nu_t$ characterizes the convergence of the global temporal tail, the spectral Rice frequency $\nu_s$ (Eq.~\ref{eq:defricefreq}), classically used in signal processing to represent expected zero-crossing rates, serves to characterize the stabilization of the average high-frequency oscillatory components of the kernel.
For a power-law kernel $G(t) \sim t^{-\alpha}$, the derivative scales as $\dot{G}(t) \sim t^{-(\alpha+1)}$.
Following a similar derivation to \cref{prop:scaling_law}, the truncation error for the spectral frequency $\Delta \nu_s(T) = \nu_s(\infty) - \nu_s(T)$ follows:
\begin{equation}
    \Delta \nu_s(T) \sim \mathcal{O}(T^{1-2\alpha})
\end{equation}
Comparing this to the temporal spread scaling $\mathcal{O}(T^{3-2\alpha})$, it is evident that $\nu_s$ converges significantly faster.  
This implies that the spectral properties stabilize significantly faster than the temporal extent of the system memory.

\section{Physical Implications and Identifiability}

\subsection{2D vs. 3D Identifiability}
The lack of a window-independent characteristic memory time in the 2D Euler limit is a specific consequence of the planarity of the wake.
It is contrasted with the 3D case to demonstrate that identifiability depends on the asymptotic decay of the vorticity field.

\begin{itemize}
    \item \textbf{2D Airfoil (Planar Wake):} The starting vortex is modeled as an infinite line vortex. 
The resulting induced velocity at the airfoil surface decays as $G(t) \sim t^{-3/2}$. 
As established in Section~\ref{sec:euler}, $M_2$ diverges logarithmically, leading to $\nu_t \sim \sqrt{\ln T}$.
\item \textbf{3D Wing (Finite Span):} For a finite-span wing, the trailing tip vortices and the starting vortex eventually coalesce into a closed loop. 
The induced velocity, in this case, decays faster than from a 2D line vortex, yielding an asymptotic impulse response typically scaling as $G(t) \sim t^{-3}$~\citep{katz2001}.
\end{itemize}

Since $\alpha > \frac{3}{2}$ for typical 3D inviscid configurations, the second moment $M_2$ converges to a finite limit. 
This implies that 3D inviscid systems admit a stable state-space representation, whereas 2D inviscid operators do not. 
The lack of a window-independent representation  identified herein is, therefore, a fundamental limitation specific to planar (or effectively 2D) formulations, including high-aspect-ratio configurations. 

\subsection{SysID Limitations}

Based on the presented results, within the inviscid limit, a clear dichotomy emerges: the characteristic memory time grows without bound, $\nu_t(T) \sim \sqrt{\ln T} \to \infty$, while the spectral Rice frequency $\nu_s(T)$ converges rapidly to a finite constant.
This indicates that local oscillatory dynamics are well-defined, whereas the global temporal spread remains unbounded.
These results imply that classical state-space models may capture local behavior but cannot represent the global temporal structure of inviscid aerodynamic responses.

Since the characteristic memory time depends on the observation window $T$, finite-dimensional state-space models may fail to capture the long-duration, low-frequency content of inviscid impulse responses.
While classical SysID seeks a fixed characteristic scale (a ``state''), the $\sqrt{\ln T}$ growth in the Euler limit implies that the effective system representation shifts continuously as the observation window grows.
Consequently, any finite-dimensional state-space approximation of an inviscid flow is not a physical property of the flow, but an artifact of the chosen observation horizon $T$.

\section{Numerical Results and Discussion}
\subsection{Computational Setup (SU2)}
To evaluate the predicted logarithmic divergence numerically, the SU2 open-source CFD suite~\citep{su2} is used to obtain the impulse response of a two-dimensional airfoil.

The simulations are performed using a symmetric NACA 0012 airfoil with a sharp trailing edge, operating in a compressible, subsonic free-stream Mach number of $M_\infty = 0.3$.
The computational domain is spatially discretized using a structured C-grid topology, using a mesh containing $419 \times 81$ grid points.
The original grid (size $449 \times 129$ grid points, extending to 500 chord-lengths in the farfield), designed for Reynolds-Averaged Navier-Stokes (RANS) simulations, was obtained from the NASA Turbulence Modeling Resource Github repository~\citep{nasatmr_github_2026, rumsey2010}.
It was coarsened in the radial direction and limited to a radial extent of $100$ chord-lengths for the present inviscid calculations.
Convective fluxes are evaluated using the L2ROE scheme~\citep{osswald2016}, coupled with the Monotonic Upwind Scheme for Conservation Laws (MUSCL) framework for second-order spatial reconstruction~\citep{vanleer1979}.
Time integration is carried out using implicit dual time-stepping with a nondimensional time step of $\Delta t = \Delta t^{*}\, U_{\infty} / c = 0.001$, where $\Delta t^*$ is the physical time step, $U_{\infty}$ is the freestream velocity, and $c$ is the chord length.

The computational procedure is designed to isolate the impulse response kernel $G(t)$.
The simulation is initialized from a steady-state solution at zero angle of attack. 
A rapid but smooth perturbation is then introduced through a plunge (heave) motion to achieve an effective $1^\circ$ change in angle of attack.

\begin{figure}
    \centering
    \includegraphics[width=0.4\textwidth]{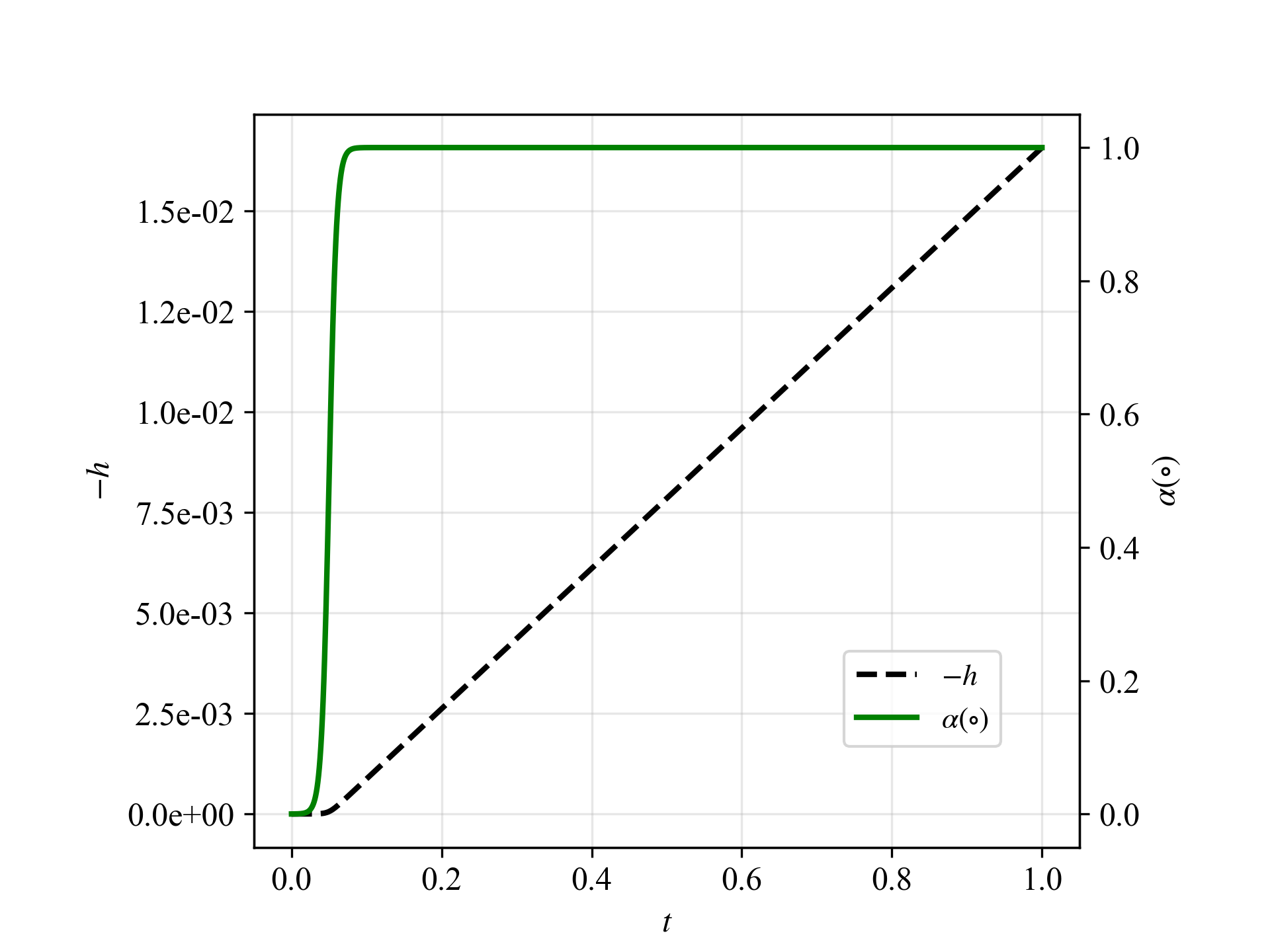}
    \caption{Vertical displacement $h(t)$ and corresponding angle of attack $\alpha(t)$ profiles in the CFD simulation, shown for the early part of the manuever ($t \le  1$).}
    \label{fig:alpha_vs_time}
\end{figure}

The imposed plunge kinematics are defined through a smoothed step in vertical velocity:
\begin{equation}
\dot{h}(t) = \frac{\alpha_{\text{AoA}}}{2} \left(1 + \tanh\left(\frac{t - t_c}{\tau}\right)\right),
\end{equation}
where $\alpha_{\text{AoA}} = \frac{\pi}{180} \text{rad} (=1^\circ$), $\tau$ controls the transition sharpness, and $t_c = 5\tau$ centers the transition.
The smoothing parameter $\tau$ is chosen such that the input approximates a step input with sufficiently sharp transition to recover the impulse response via differentiation.
This setup provides a direct numerical realization of the system’s indicial response, from which the underlying impulse response can be extracted.
The profiles of the vertical displacement, $h(t)$ (nondimensionalized by $c$), and instantaneous angle of attack, $\alpha(t) (=-180/\pi\dot{h}(t))$, are shown in Figure~\ref{fig:alpha_vs_time} during the early part of the maneuver.

The aerodynamic impulse response kernel is obtained from the time history of the lift coefficient.
Specifically, the kernel is defined as:

\begin{equation}
G(t) = \frac{d C_L}{dt}.
\end{equation}

To compute this derivative, a Savitzky–Golay filter is applied to the lift signal.
This approach ensures sufficient smoothness for differentiation while preserving the physically relevant high-frequency transients present in the early-time response.

\subsection{Models Used for Comparison}

The inviscid Euler results obtained from SU2 are compared against two reference models:

\begin{itemize}
\item \textbf{Wagner’s Approximation:}
A reduced-order representation of unsteady lift based on a sum of exponential terms. This model corresponds to a finite-dimensional, Markovian system with inherently bounded temporal moments.

\item \textbf{Analytical Euler:}
The asymptotic inviscid kernel $G(t) \sim t^{-3/2}$ is used as a theoretical reference. To avoid the singularity at $t=0$, it is evaluated in a regularized form $G(t) = (t + \epsilon)^{-3/2}$ with a small cutoff $\epsilon = 0.01$. This modification affects only the early-time behavior and does not alter the asymptotic scaling governing the temporal moments.

 \end{itemize}

These models provide two limiting cases: exponential decay, corresponding to finite-memory systems, and power-law decay at the critical rate, corresponding to marginal, non-convergent memory.

\subsection{Numerical Results}
\subsubsection{Kernel Decay}

\begin{figure}[htpb]
    \centering
    \includegraphics[width=0.5\textwidth]{fig/plotA.png}
    \caption{Impulse response kernel decay on log-log scale, comparing the analytical Euler model, CFD Euler simulation, and Wagner exponential approximation.}
    \label{fig:plotA}
\end{figure}

As shown in Figure~\ref{fig:plotA}, analysis of the impulse response kernel confirms that the analytical Euler asymptotic model follows a straight-line decay on a log-log plot with a slope of $-3/2$. 
In contrast, the Wagner approximation model exhibits a downward curve that accelerates prominently from $T=40$ onwards, indicative of its long-term exponential cutoff.
The CFD-derived Euler kernel closely tracks the overall analytical $t^{-3/2}$ decay for the majority of the observation window.
It also exhibits localized, non-monotonic deviations in the form of sharp drops and recoveries.
During the early-time response, the two prominent observed spikes in the impulse response correspond to distinct physical mechanisms.
The first spike (near $T \sim 0.07$) is attributed to added-mass effects due to the initial rapid acceleration ($\ddot{h}$) of the plunging maneuver.
The second spike (near $T \sim 0.5$) is driven by wake interaction effects, wherein the initially shed vorticity convects downstream and feeds back onto the airfoil, briefly redistributing the surface pressure loading.
At late times ($T > 100$), the CFD curve exhibits high-frequency oscillations due to infinitesimal $C_L$ variations as the limit of the numerical noise floor is reached.
The perturbations described above remain confined to short temporal scales and do not alter the global power-law decay of the kernel.
The large-scale envelope continues to follow the theoretical $t^{-3/2}$ scaling until late-time numerical dissipation induces an exponential cutoff.

\subsubsection{Characteristic Memory Time Divergence}
\begin{figure}[htpb]
    \centering
    \includegraphics[width=0.5\textwidth]{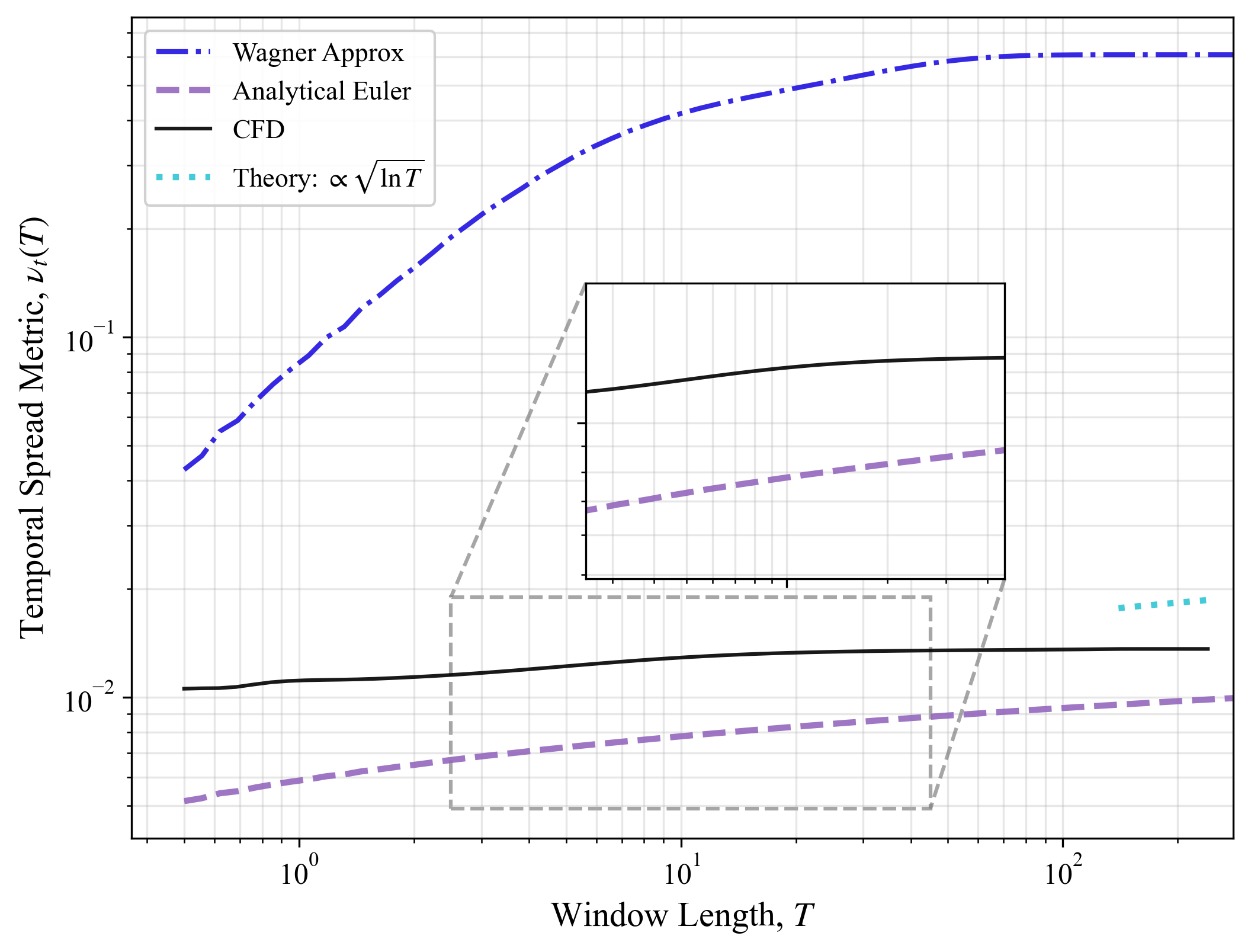}
    \caption{Temporal spread scale $\nu_t(T)$ comparing CFD-derived results with the stable plateau of the Wagner approximation and the logarithmic divergence of the analytical Euler limit.}
    \label{fig:plotB}
\end{figure}

As shown in Figure~\ref{fig:plotB}, the temporal spread scale or characteristic memory time $\nu_t(T)$ for the Wagner model rises rapidly before exhibiting a two-stage plateau. 
This is consistent with the multi-exponential structure embedded in the R.T. Jones approximation.
However, the analytical Euler curve exhibits a persistent upward drift consistent with the theoretical $\sqrt{\ln T}$ scaling.
The CFD Euler data initially tracks this theoretical gradual rise but begins to plateau much earlier, around $T=50$.
This premature late-time flattening indicates that the effective CFD system has transitioned from algebraic (power law) to exponentially damped behavior, confirming that numerical dissipation enforces an artificial finite memory.

\subsubsection{Numerical Dissipation and the Scaling Exponent}
\begin{figure}[htpb]
    \centering
    \includegraphics[width=0.5\textwidth]{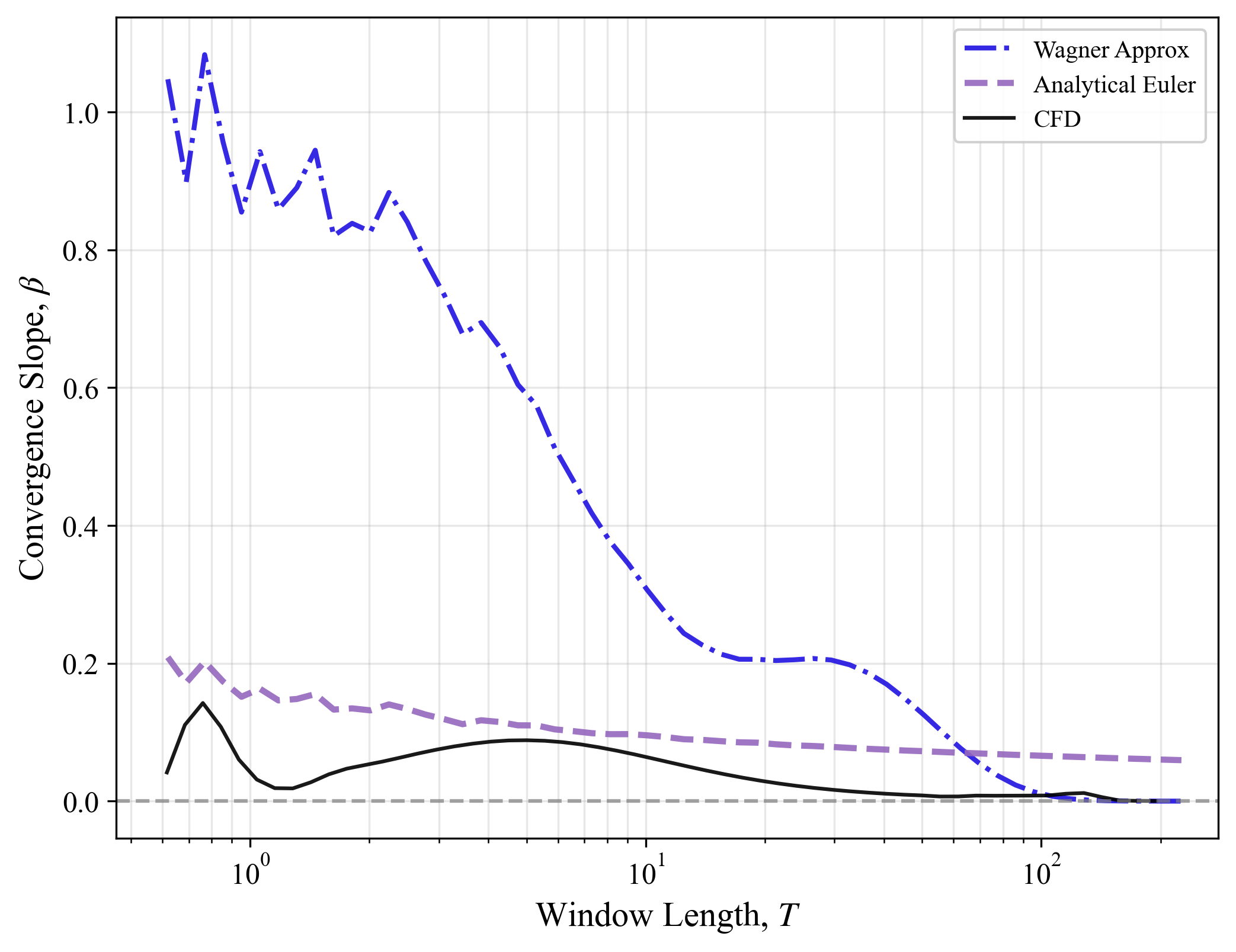}
    \caption{Effective scaling exponent $\beta_{\text{eff}}$ characterizing the growth rate of the memory time.}
    \label{fig:plotC}
\end{figure}

To map the transition where the CFD data separates from analytical theory, the effective scaling exponent is defined, $\beta_{\text{eff}} = \frac{d \ln \nu_t}{d \ln T}$, which characterizes the localized growth rate of the characteristic memory time. 
$\beta_{\text{eff}}$ theoretically scales as $1/(2 \ln T)$, and $\beta_{\text{eff}} \to 0$ does not imply the convergence of $\nu_t$; it only indicates that the growth rate approaches zero, while the function itself continues to grow without bound.

As illustrated in Figure~\ref{fig:plotC}, the analytical Euler limit exhibits the theoretical gradual drop in $\beta_{\text{eff}}$. 
By contrast, the finite-memory Wagner model rapidly drops toward zero in staged steps corresponding to its discrete time constants.
This eventual decay of $\beta_{\text{eff}}$ for CFD marks the onset of dissipation-dominated behavior.
The key observation is that the CFD Euler system converges while the theoretical Euler limit does not, confirming that convergence of temporal moments requires a dissipative mechanism, provided here by the numerical viscosity inherent to the numerical scheme used.

\subsubsection{Spectral vs Temporal Metrics}

\begin{figure}[htpb]
    \centering
    \includegraphics[width=0.5\textwidth]{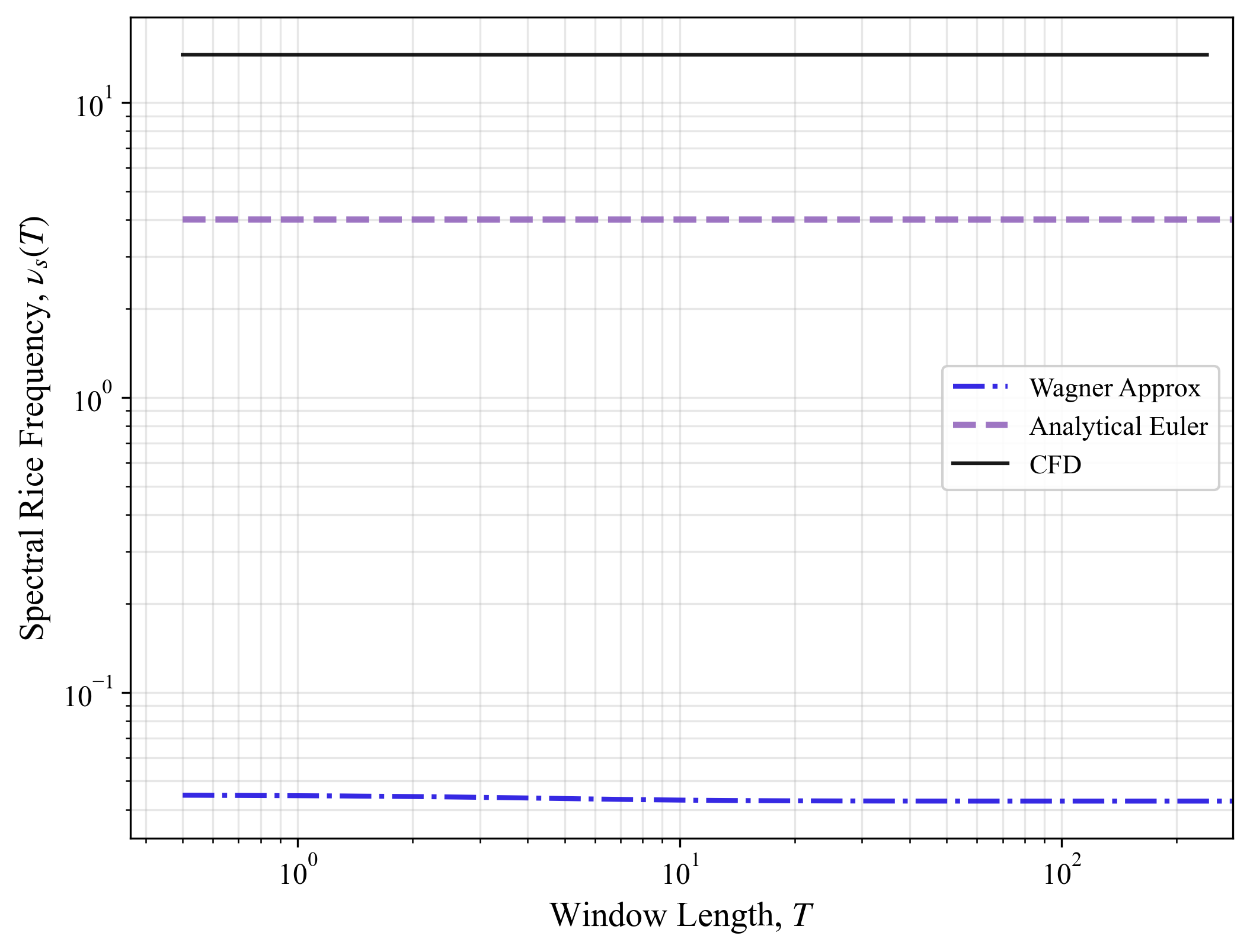}
    \caption{Classical Spectral Rice Frequency $\nu_s(T)$ showing rapid convergence for all models.}
    \label{fig:plotD}
\end{figure}

The contrasting behavior of the spectral Rice frequency, $\nu_s(T)$, is examined next.
As shown in Figure~\ref{fig:plotD}, $\nu_s(T)$ exhibits rapid convergence, with all three curves remaining flat throughout the observation window.
The distinct magnitudes at which they settle reflect the initial high-frequency smoothness of each kernel.
This behavior demonstrates that the local oscillatory characteristics of the inviscid kernel (high-frequency content) are physically well-defined and effectively invariant over the observation window.
Conversely, the characteristic memory time $\nu_t(T)$ shows no such internal stabilization. 

\subsection{Implications for System Identification}
The transition from theoretical continuous divergence to numerical plateau underscores a fundamental principle: identifiability of an aerodynamic system via temporal moment-based representations requires a dissipative mechanism.

In CFD simulations, numerical dissipation (arising from spatial discretization, artificial viscosity, and time-stepping schemes) acts as an effective regularizer that enforces exponential decay at long times.
In physical systems, molecular viscosity fulfills this role, dissipating the kinetic energy of shed vorticity and ensuring that the impulse response exhibits sufficiently fast decay.

The artificial memory plateau observed in the CFD data should not be mistaken for proof of the analytical Euler's asymptotic behavior; it is a signature of dissipative regularization dominating the late-time regime.
Consequently, when a stable finite-dimensional SysID model successfully converges on CFD Euler data, the resulting long-horizon states do not represent pure inviscid fluid physics; they are parameterizing the effective dissipation introduced by the numerical method.

For reduced-order modeling, the consequence is as follows: classical state-space identification applied to inviscid flow data yields models that are artifacts of the observation window and the regularization mechanism (numerical or physical), rather than intrinsic properties of the flow.

\section{Conclusion}
This study investigates the compatibility of finite-dimensional state-space models with the asymptotic memory structure of two-dimensional inviscid unsteady aerodynamics.
By formulating a temporal-moment diagnostic $\nu_t(T)$ describing the characteristic memory time scale, the $t^{-3/2}$ power law decay of the inviscid Euler impulse response is mathematically proven to lie exactly on the critical threshold for moment convergence.
At this boundary, the second temporal moment diverges logarithmically, causing the effective memory time to grow as $\sqrt{\ln T}$ with the observation window.
Therefore, 2D scale-free aerodynamic systems do not admit a window-independent temporal memory scale.
This behavior contrasts with the rapid convergence of the spectral Rice frequency, indicating that while local oscillatory content is physically well-defined, global temporal spread remains unbounded.
The compressible Euler CFD simulation of a 2D plunging airfoil is compared with these theoretical findings.
The computed impulse response effectively tracked the theoretically predicted logarithmic growth at intermediate time horizons.
However, at late times, numerical dissipation inherent to discretization imposed an artificial cutoff, producing a stable plateau consistent with fading-memory behavior.
This work establishes that any finite-dimensional state-space model fitted to 2D inviscid CFD data effectively parameterizes the observation horizon, rather than providing a window-independent characterization of the intrinsic flow memory.

\newpage
\bibliographystyle{plainnat}
\bibliography{main}

\end{document}